\documentclass[preprint,10pt,a4paper]{aastex}

\usepackage{graphicx}
\usepackage{epstopdf}
\usepackage{bm}
\usepackage{lineno}

\newcommand{\be}{\begin{equation}}
\newcommand{\ee}{\end{equation}}

\newcommand{\ba}{\begin{eqnarray}}
\newcommand{\ea}{\end{eqnarray}}
\newcommand{\om}{\omega}

\newcommand{\Bf}{{magnetic field}}

\newcommand{\NS}{neutron star}
\newcommand{\ms}{magnetosphere}
\newcommand{\mss}{magnetospheres}
\newcommand{\NSs}{{neutron stars}}

\newcommand\eg{{\it{e.g.}}}

\newcommand\lo{\mathrel{\raise.3ex\hbox{$<$}\mkern-14mu\lower0.6ex\hbox{$\sim$}}}
\newcommand\go{\mathrel{\raise.3ex\hbox{$>$}\mkern-14mu\lower0.6ex\hbox{$\sim$}}}

\begin{document}
%\runauthor{Lyutikov}
%\begin{frontmatter}
\title{Radius-to-frequency mapping 
%in  \mss\ of  \NSs\ 
and FRB frequency drifts}

\author{
Maxim Lyutikov,
Department of Physics and Astronomy, Purdue University, 525 Northwestern Avenue, West Lafayette, IN, USA 47907}

\begin{abstract}
We build a model of   radius-to-frequency mapping in  \mss\ of  \NSs\ and apply it to frequency drifts observed in Fast Radio Bursts. We assume that an emission patch propagates along the dipolar \Bf\ lines producing coherent emission   with frequency, direction and polarization  defined by the local \Bf.  The observed temporal evolution of the frequency depends on relativistic effects of time contraction and the curvature of the magnetic field lines.
The model generically produces linear scaling of the drift rate, $\dot{\om} \propto - \om$, matching both numerically and parametrically the rates observed in FBRs; a  more complicated behavior of  $\dot{\om}$ is also possible. Fast rotating \mss\  produce higher drifts rates for similar viewing parameters than the slowly rotating ones.
In the  case of repeaters same source  may show variable drift pattens depending on the observing phase. We expect rotational of polarization position angle through a burst, though by  smaller amount than in radio pulsars. All these findings compare favorably with properties of FBRs, strengthening their possible {\it loci}  in the  \mss\ of  \NSs.
 \end{abstract}

\section{Introduction}

Fast Radio Bursts (FRBs) \citep{2007Sci...318..777L,2019A&ARv..27....4P,2019arXiv190605878C} is a recently identified enigmatic  astrophysical  phenomenon. A particular sub-class of FRBs -  the repeating FRBs -  show similar downward drifting features in their dynamic spectra: FRB121102
\citep{2019ApJ...876L..23H}, FRB180814 \citep{2019Natur.566..235C}, and lately numerous FRBs detected by CHIME \citep{2019arXiv190803507T,2019arXiv190611305J}. 
The properties of the drifting features are highly important for the identification of the {\it loci} of FRBs, as discussed by \cite{2019arXiv190807313L}.

First,  generation of narrow spectral features is  natural in the ``plasma laser" concept of coherent emission generation, either due to the discreteness of plasma normal modes related to the spatially  local plasma parameters (\eg\ plasma and cyclotron frequencies) or changing resonant conditions. Frequency drift then reflects the propagation of the emitting particles in changing magnetospheric conditions, similar to what is called ``radius-to-frequency mapping'' in pulsar research \citep[\eg][]{1977puls.book.....M,1992ApJ...385..282P}.

Second,  drift rates  and  their frequency scaling can be used to infer the physical size of the emitting region  \citep{2019arXiv190807313L}. \cite{2019arXiv190611305J} \citep[see also][]{2019ApJ...876L..23H} cite a drift rate for FRB 121102 of 
\be
\partial_t \ln \om \approx -150  s^{-1}
\label{drift00}
\ee
extending for  an order of magnitude in frequency range. 
This implies that: (i) emission properties are self-similar (\eg\ power-law scaled); (ii)  typical size
\be
\Delta r \sim \frac{c}{  \partial_t \ln  \om} = 2 \times 10^8 {\rm cm}
\ee
Both these estimates are consistent with emission been produced in \mss\ of \NS. 

In this paper we build a model of radius-to-frequency mapping for (coherent) emission generated by relativistically moving sources in \mss\ of \NS.  The concept of 
``radius-to-frequency mapping''  originates in pulsar research \citep[\eg][]{1977puls.book.....M,1992ApJ...385..282P}. The underlying assumption is that at a given place in the \mss\ of pulsars the plasma produces emission specified by the local, radius-dependent properties. This general concept {\it does not specify}  a particular emission mechanism, just  assumes that the properties are radius-dependent. 

As a working model we accept the ``magnetar radio emission paradigm", whereby the coherent emission is magnetically powered, similar to solar flares, as opposed to rotationally powered in the case of pulsars \citep{lyutikovradiomagnetar,2013arXiv1307.4924P}. 
Recently, \cite{2019ApJ...882L...9M} discussed how many properties of magnetar radio emission resemble those of FRBs (except  the frequency drifts, see \S \ref{driftrotate})

 Rotationally-powered FRB emission mechanisms \citep[\eg\ as analogues of Crab giant pulses][]{2016MNRAS.462..941L} are excluded by the  localization of the Repeating FRB at $\sim 1$ Gpc  \citep{2016Natur.531..202S}, as discussed by \cite{2017ApJ...838L..13L}. Magnetically-powered emission has some observational constraints, but remains theoretically viable  \citep{2019arXiv190103260L,2019arXiv190807313L}.

Within  the ``magnetar radio emission paradigm", the coherent emission is generated on closed field lines, presumably due to reconnection events in the magnetosphere. The observed properties then depend on: (i) particular scaling of the emitted frequency  $\om$ on the  emission radius  $r_{em}$ --  $\om(r_{em})$ --  we leave this dependence unspecified;  (ii)
motion of the emitter  -- we assume  motion along the \Bf\ line;  (iii) emission beam -- we assume  that emission is  along the local  \Bf\ lines; (iv) line of sight through the  spinning \ms.  

In this paper we consider all the above effects. First, in \S \ref{stat} we consider  stationary \mss, and then in \S \ref{rot} the spinning ones.

\section{ Emission kinematics with relativistic  and curvature effects}
\label{1}

\subsection{Model set-up}

An important concept  is the observer time - a time measured from the arrival of the first emitted signal \citep[see, \eg\  models of Gamma Ray Bursts,][]{2004RvMP...76.1143P}.  In our case both the relativistic motion and  the curved  trajectory  strongly affect the relation between the coordinate time $t$ and  the  observer time $t_{ob}$.  To separate effects of rotation from the propagation, we first consider stationary \mss.

\begin{figure}[h!]
  \centering
  \includegraphics[width=0.99\textwidth]{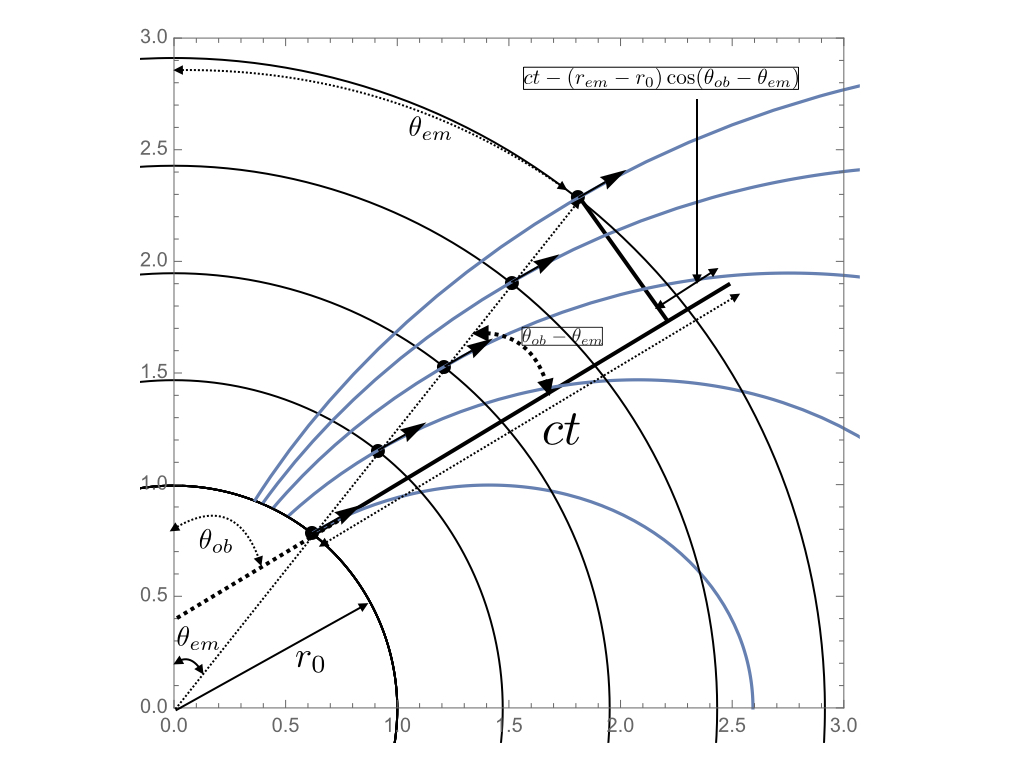}
  \caption{Location of emission points within the  \ms.  Magnetic axis is vertical. Observer is located at  polar angle $\theta_{ob}$, which is time-dependent for the rotating case.  At time $t=0$ an emission front is lightened from the surface $r_0=1$, propagating along the local \Bf\ lines with velocity $\beta$.   Circles correspond to the radius $r_{em}$ of the emission points at times $t=0, 0.5, 1, 2$;  emission points at each moment are located at  radius $r_{em}$  and polar angle $\theta_{em}$; emission is along the local \Bf .   Due to the field lines curvature the emission front at later times  lags behind the one emitted at $t=0$ even for $\beta=1$.   Observer angle $\theta _{ob}=\pi/4$  in the example pictured. The insert indicates the relation between the observer time and geometrical  parameters.
 }
  \label{movingout}
  %file FRB-drift
 \end{figure}

Let's assume that 
at time $t=0$ an  emission front is launched from radius $r_0$, propagating with velocity $\beta c$  {\it along} the local \Bf, Fig. \ref{movingout}.   Thus, we assume that {the whole of the \ms\ starts to produce emission instantaneously.}  The trigger could be, \eg\ an onset of magnetospheric reconnection event \citep{2006MNRAS.367.1594L, 2015MNRAS.447.1407L}. A reconnection event that encompasses the whole  region near the surface of the \NS\ will be seen at some distance away as a coherent large scale event.   
    If only a patch of the \ms\ produces an emission, the light curves will be truncated  accordingly. Given the fact that we already have a number of model  parameters, we did not explore the finite size of the emission regions in the $r-\theta-\phi$ space.

 Let the observer be  located at an angle $\theta_{ob}$, measured from the instantaneous direction of the magnetic dipole. Angle  $\theta_{ob}$ defines a field line with a tangent (the \Bf) along the line of sight at the radius $r_0$. That  field line can be  defined by the angle $\theta_0$ of the magnetic foot point.  At time $t$ the photons emitted from the surface  at $t=0$ propagated a distance $c t$.  It is assumed that at each point  emission is produced  along the line of sight  (solid points and arrows in Fig. \ref{movingout}). We assume that at given location $r_{em}$ the emission front produces coherent emission at a frequency $\om(r_{em})$. (We neglect the fact that in a dipolar \ms\ the strength of the  \Bf\ at given radius varies by a factor of $2$ depending on the magnetic latitude.) As the emission front propagates in the \ms\ different \Bf\ lines contribute to the observed emission.  At time $t$  emission point is located at distance $r_{em}$ and the polar angle $\theta_{em}$. Emission from the point $r_{em}$ arrives at the observer at time $t_{ob}$ that depends on: (i) emission time;
 (ii) velocity of the emission front; (iii) geometry of field lines.

For a field line parametrized by polar angle $\theta_0$ at $r_0$,  a distance along the field line from $\theta_0$ to $\theta> \theta_0$ is
\be
%\ba &&
\Delta s = \frac{r_0}{\sin^2 \theta_0}  \int_{\theta_0} ^ \theta \sqrt{1+3 \cos ^2 \theta}  \sin \theta  d \theta
%\nn &&
 %\frac{r_0}{\sin^2 \theta_0}\frac{1}{6} \left(-3 \cos (\theta ) \sqrt{3 \cos ^2(\theta )+1}+3 \cos \left(\theta   _0\right) \sqrt{3 \cos ^2\left(\theta _0\right)+1}-\sqrt{3} \sinh ^{-1}\left(\sqrt{3}   \cos (\theta )\right)+\sqrt{3} \sinh ^{-1}\left(\sqrt{3} \cos \left(\theta   _0\right)\right)\right)
 %  \ea
 \label{DS0}
 \ee
 
 Emitting particles move according to 
 \be
 \Delta s  = \beta t
 \label{DS}
 \ee
 where $t$ is the coordinate time. 
 %(For $\beta =1$ the time it takes to reach magnetic equator starting at a foot-point $\{r_0,\theta_0\}$ is larger than $r_0/c$ for $\theta_0 \leq 0.70$.   This corresponds to $24\%$ of the total area.)

The points $\theta_{em}$ in the dipolar \ms\ that have \Bf\ along the line of sight satisfy
\be
\cos 2 \theta_{em} = 
\frac{1}{6} \left(\sqrt{2}  \cos\theta _{ob} \sqrt{ \cos
   \left(2 \theta _{ob}\right)+17}+\cos \left(2 \theta
   _{ob}\right)-1\right)
   \label{thetam}
   \ee
   
   As  figure \ref{movingout} demonstrates, the observer time  is given by (speed of light is set to unity)
 \be
 t_{ob} = t -( r_{em} -r_0)\cos   \left( \theta_{ob}- \theta_{em} \right)
 \label{tob}
 \ee
with $r_{em}(t)$ given by the requirement that particles propagating along the curved field with velocity $\beta$ emit along the local \Bf. 
For nearly straight field lines and highly relativistic velocity,  $\beta \approx 1-1/ (2  \Gamma^2)$, the effects of field line curvature dominate for 
$\theta_{ob}- \theta_{em} \geq 1/\Gamma$. 
  (In calculations below the  time is normalized to  unites $r_0/c$, where $r_0$ is some initial radius, not necessarily the \NS\ radius.)
%(For example, for straight field lines  $\theta_{ob}= \theta_{em}$, $r_{em}=1+\beta t$,  Eq. (\ref{tob}) gives $t_{ob} = t/ (2 \Gamma^2)$ for $\beta \approx 1-1/ (2  \Gamma^2)$.

 We then implement the following procedure, Fig. \ref{movingout}
   \begin{itemize}
    \item Given is  the observer angle  $\theta _{ob} $ (with respect to the magnetic dipole). 
    \item Find the polar angle of the footprint  $\theta_0 ^{(0)} $ by solving Eq.  (\ref{thetam}) and setting $\theta_{em}=\theta_0 ^{(0)}$. (Superscript $^{(0)} $ indicates the moment $t=0$.
    \item After time $t$ the emission front moved along the field lines according to Eqns (\ref{DS0}-\ref{DS}), where $\theta_0(t)$ is a parameter for the field line emitting at time $t$ (at $t=0$ we have  $\theta_0(0)= \theta_0 ^{(0)}$)
   \item For $t\geq 0$, using Eqns.  (\ref{DS0}-\ref{DS}-\ref{thetam}) with $\theta=\theta_{em}$ ,   find the polar angle of the foot-point  $\theta_0$,  (\ref{thetam}), where \Bf\ is along the line of sight at time $t$.
    \item Using  (\ref{DS}) find $\theta_0$ - the polar angle of the field line that produces emission at time $t$.
    \item Given time $t$ and the location of the emission point we can calculate the observer time, Eq. (\ref{tob}).
    \item We then find dependence of  $ r_{em}$ versus $t_{ob}$.
    \item Assuming some $\om (r_{em})$ we find the radius-to-frequency mapping  $\om (t_{ob})$.
    \end{itemize}

    Thus, we take into account relativistic transformations and curvature of the field lines in calculating the relations between the  observer time versus the coordinate time.
    We implicitly assume that emitted frequency is the function of the emission radius, $\om ( r_{em})$, but given our uncertainty about the emission properties we do not specify a particulate dependence $\om ( r_{em})$. We plot curves for generic profiles $ \om  \propto r_{em}^{-1}$ and  $ \om  \propto r_{em}^{-3}$; the last scaling is expected  if the emission is linearly related to the local \Bf.

   % We assume that emission continues for $0<t<1$.
    
    The velocity of the emitting front has a complicated effect on the overall duration of the observed pulse and a range of emitted frequencies. For sub-relativistic velocities 
    $\Delta r_{em} \sim \beta \Delta t$ and $t_{ob} \sim  \Delta t$. As $\beta \rightarrow 1$, the observed duration shortens,  $t_{ob} \ll  \Delta t$. But for sufficiently high velocity, $\beta \approx 1$ with $\theta_{em} -\theta_{ob} \geq 1/\Gamma$,  this relativistic line-of-sight contraction become unimportant, as the observed duration is determined by the curvature of field lines.

\subsection{Results: stationary \ms}
\label{stat}

\begin{figure}[h!]
  \centering
  \includegraphics[width=0.8\textwidth]{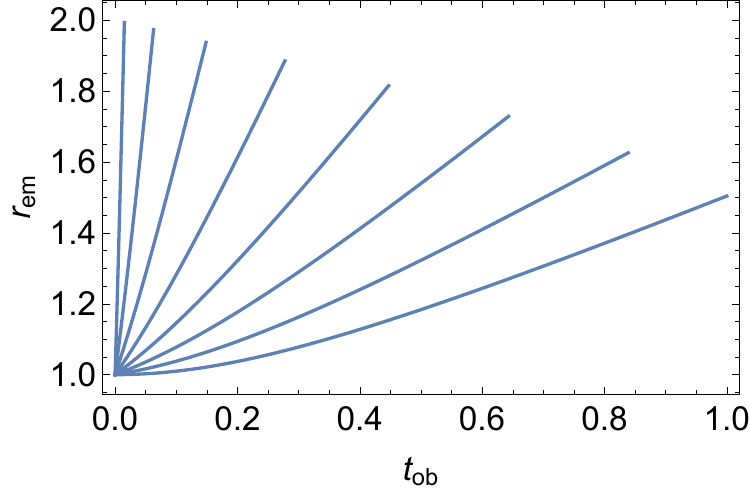}
  \caption{Emission radius as a function of the observer time for different viewing angles, $\theta_{ob} = \pi/8, \,\pi/4 ... \pi$; non-rotating \ms; $\beta=1$, duration of  propagation in coordinate time  is $\Delta t=1$. At larger viewing angles the field lines are more curved - this cancels the relativistic line-of-sight effects,  producing longer duration pulses even for $\beta =1$. (For $\beta=1$ and  $\theta_{ob}=0$ all emission arrives at $t_{ob} =0$).  For angles   $\theta_{ob} > \pi/2$ it is assumed that only ``upper'' half of the \ms\ emits.}
  \label{FBR-drift}
  %file FRB-drift
 \end{figure}

In Fig. \ref{FBR-drift} we implement the above-describe procedure showing $ r_{em}(t_{ob})$ for the extreme relativistic limit of  $\beta=1$. The figure  demonstrates that the effects of \Bf\ line curvature can dominate over the relativistic effects along the line of sight.  

\begin{figure}[h!]
  \centering
  \includegraphics[width=0.3\textwidth]{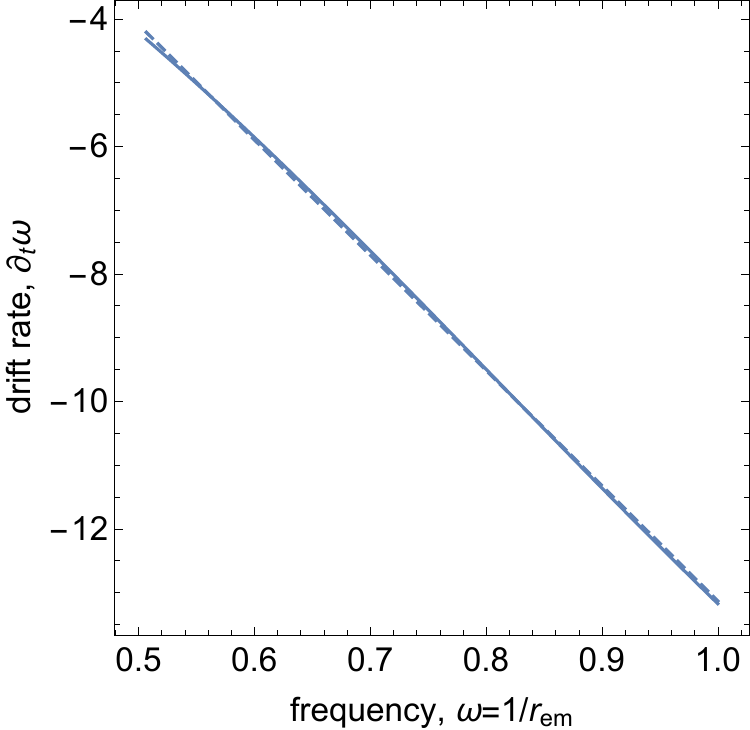}
  \includegraphics[width=0.3\textwidth]{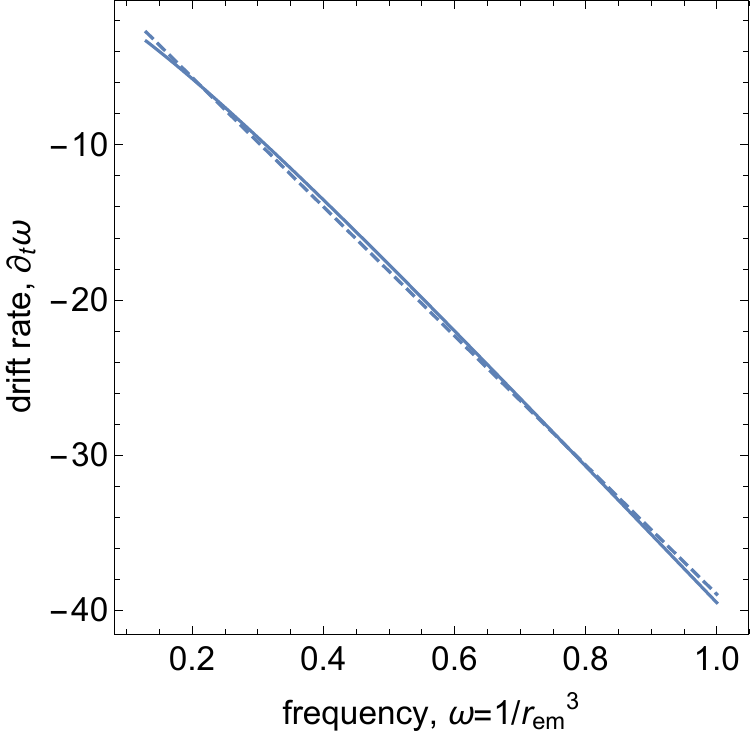}\\
  \includegraphics[width=0.3\textwidth]{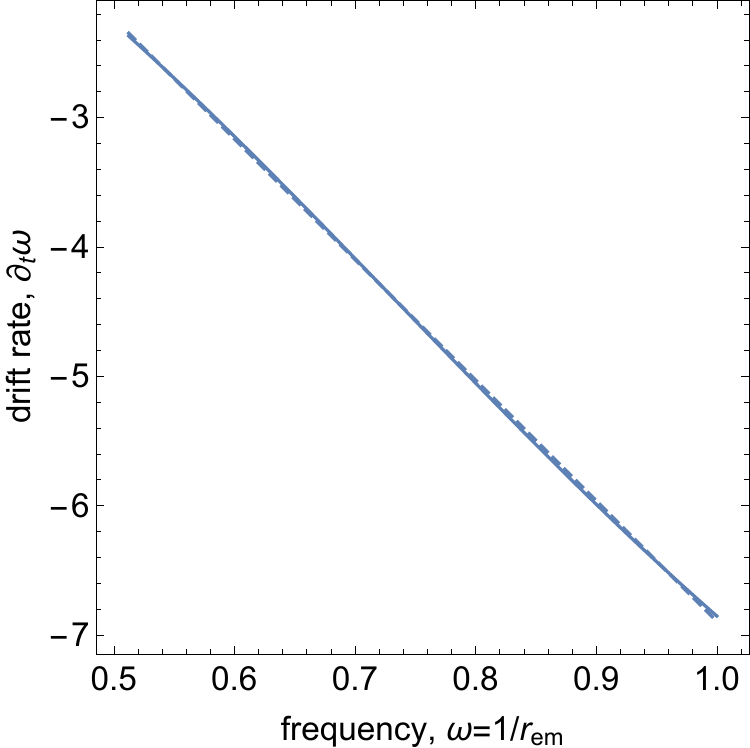}
  \includegraphics[width=0.3\textwidth]{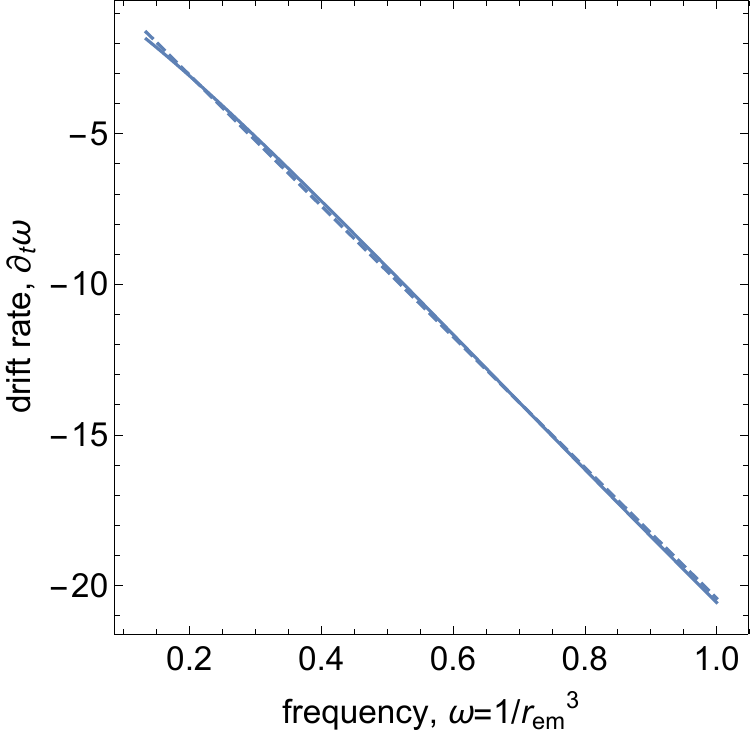}\\
   \includegraphics[width=0.3\textwidth]{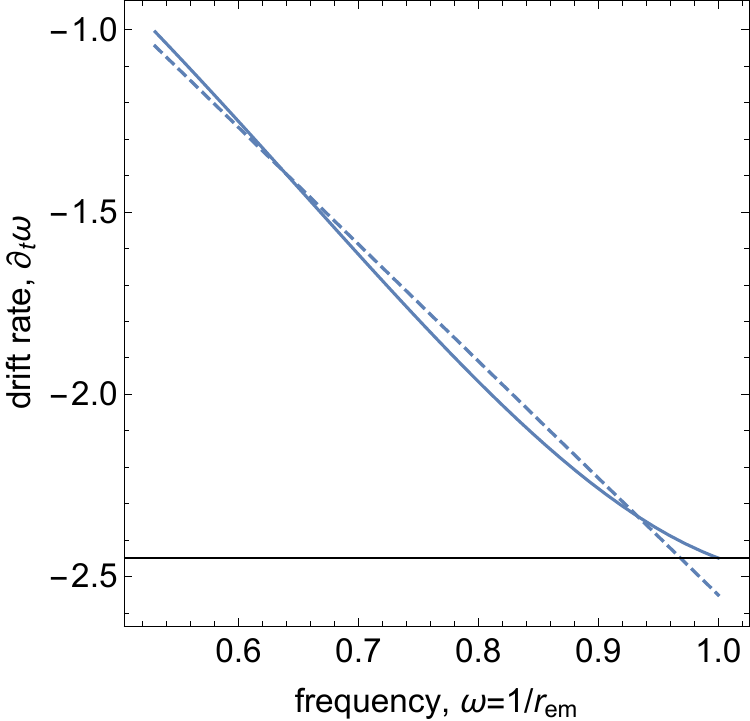}
  \includegraphics[width=0.3\textwidth]{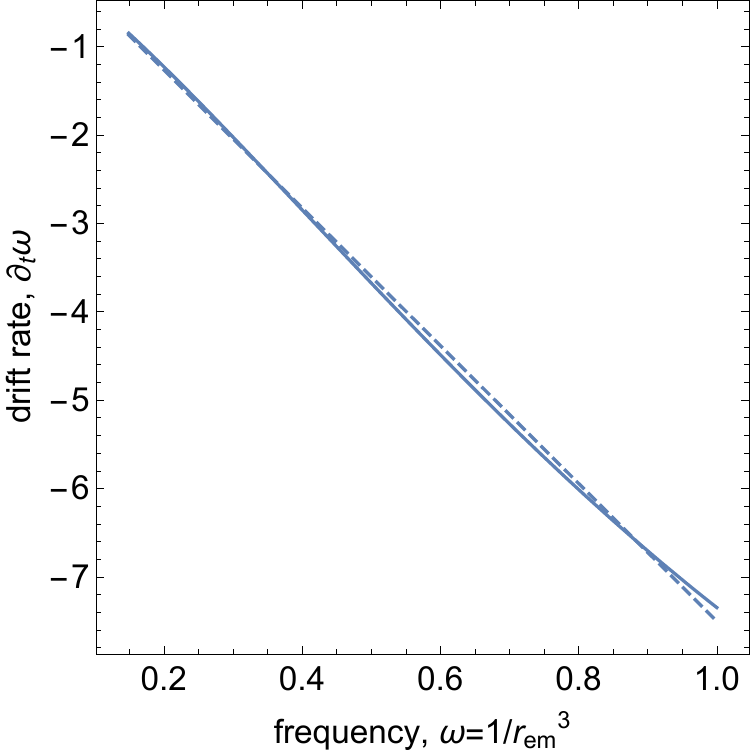}
  \caption{Drift rates in a stationary \ms\ as function of frequency for two scalings: $\om \propto r_{em}^{-1}$ (left panels)  and  $\om \propto r_{em}^{-3}$ (right panels). Non-Rotating \mss. Top row: $\theta _{ob} =\pi/4$, middle row  $\theta _{ob} =\pi/3$, bottom row  $\theta _{ob}=\pi/2$.  Dashed lines are linear fits $\dot{\om}  \propto \om$.  This simplest case demonstrates that for most observer angles the frequency drift is linear in time (for smaller $\theta _{ob}$ the drift is more linear since the fields lines are straighter  near the magnetic pole.}
  \label{drift11}
  %file FRB-drift
 \end{figure}

For assumed scaling $\om(r_{em})$  $\om \propto r_{em}^{-1}$ and  $\om \propto r_{em}^{-3}$ the corresponding curves $\om(t_{ob})$ are given in Fig. \ref{drift11}.
 The model  generally reproduces, approximately linear drifts rates \citep[see, eg][Fig. 6]{2019arXiv190611305J} regardless of the particular power law dependence   $\om(r_{em})$.     We consider this as a major success of the model.

   \subsection{Rotating \ms}
   \label{rot} 
   Assume next that the star  is rotating with spin frequency $\Omega$. The magnetic polar angle of the line of sight at time $t$ {\it  in the pulsar frame }   is then 
 \be
 \cos \theta _{ob}=\cos\alpha  \cos\theta
   _{ob}^{(0)}+\sin\alpha  \sin \theta _{ob}^{(0)} \cos
   (\Delta \phi +t \Omega )
   \label{2}
   \ee
   where $\alpha$ is the inclination angle between rotational axis and magnetic moment, $\theta _{ob}^{(0)} $ is the observer angle in the plane comprising 
   vectors of $\Omega$, $\mu$ and the line of sight, and $\Delta \phi$ is the azimuthal angle of the observer with respect to the $\Omega$-$\mu$ plane when the injection starts - this is the phase at $t=0$.
   (For example, if emission starts when the line of sight is in the $\Omega$-$\mu$ plane, at that moment $\theta _{ob}= \alpha - \theta _{ob}^{(0)}$.)
   
     \begin{figure}[h!]
  \centering
  \includegraphics[width=0.8\textwidth]{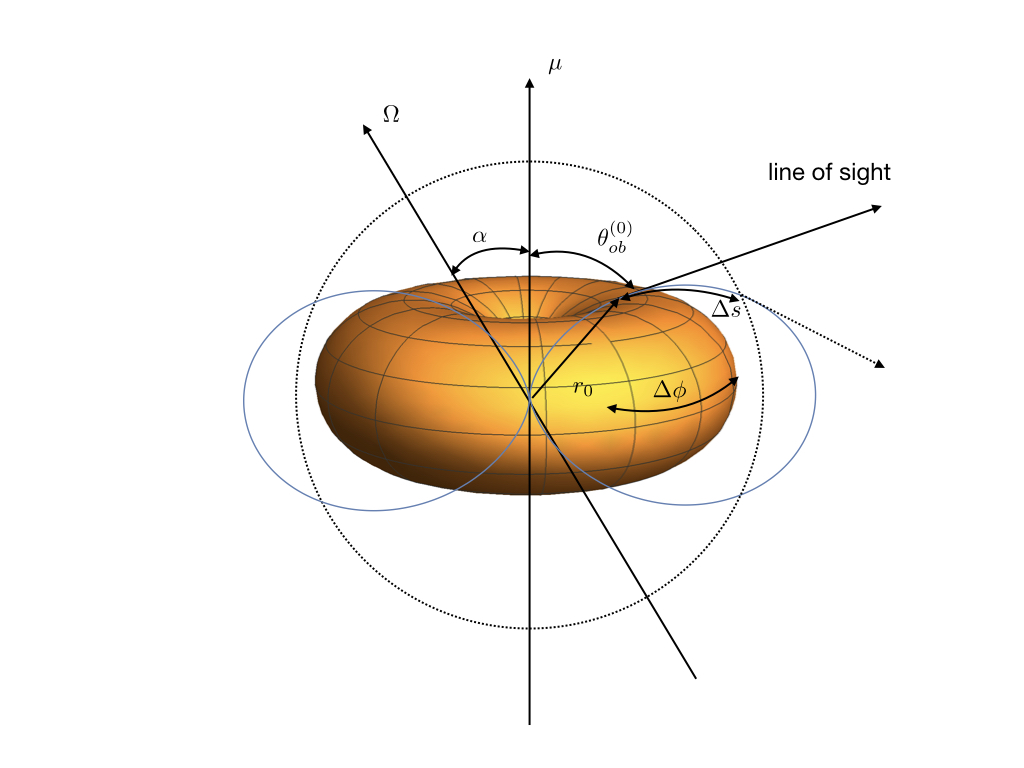}
  \caption{ Geometry of the problem at the moment when the line of sight is in the $\mu - \Omega$ plane; reference frame associated with the \NS. Magnetic moment is inclined by the angle $\alpha$ with respect to the rotation axis $\Omega$. When the line of sight is in the  $\mu - \Omega$ plane, the angle between the line of sight and magnetic moment is   $\theta _{ob}^{(0)}$. Emission starts at $r_0$ (at the moment defined by $\Delta \phi$ and propagates along the local \Bf\ according to $\Delta s = \beta t$. Emission is along the local \Bf. Later times are denoted by dotted lines. The model is inherently 3D - this picture only illustrate the main geometrical factors.}
   \label{FBR-drift-geom}
  %file FRB-drift
 \end{figure}

   We then implement the  procedure, outlined in \S \ref {1}, with the following modifications, see Fig. \ref{FBR-drift-geom}
   \begin{itemize}
    \item Given are the $\theta _{ob}^{(0)} $, $\alpha$, $\beta$, $\Omega$ and $\Delta \phi$
   \item  For $t \geq 0$ implement procedure of  \S \ref {1} with time-dependent  $\theta _{ob}$ given by Eq. (\ref{2})
    \end{itemize}

      \subsubsection{Frequency  drifts in rotating \ms}
\label{driftrotate}
In the rotating \mss\ the observed frequency drifts are generally more complicated, as the line of sight samples larger part of the \ms. A key limitation in the approach is that we assume that the whole surface $r=0$ produced as emission front - thus different parts of the emission front maybe casually disconnected - under certain circumstances  this leads to unphysical results (\eg\ upward frequency drifts). 
   
\begin{figure}[h!]
  \centering
  \includegraphics[width=0.8\textwidth]{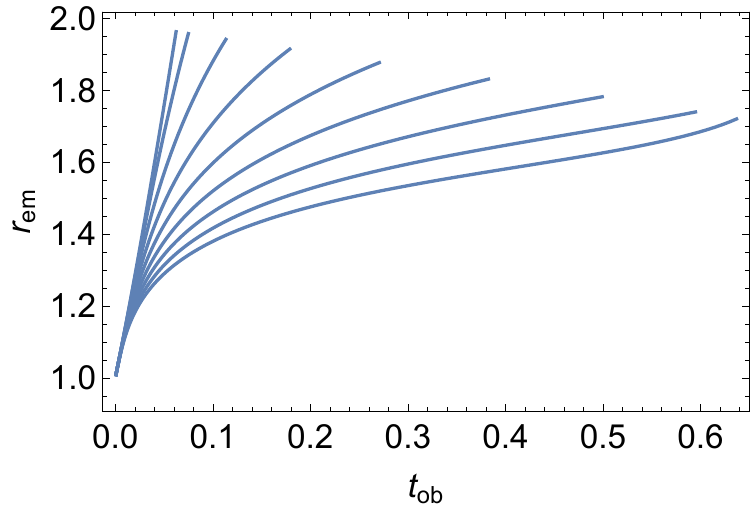}
  \caption{ Emission radius as function of the observer time for different $\Omega = 0, \pi/8... \pi$ (left to right, $\alpha = \pi/4$, $\delta \phi = 0$, $\theta_{ob} = \pi/2$). This plot demonstrates that faster spin  increases the observed duration of a pulse, as the line of sight samples larger  parameter space.}
   \label{remofTobOmega}
  %file FRB-drift
 \end{figure}

   In Fig. \ref{remofTobOmega} we plot emission radius as function fo observer time for different pulsar spins. It is clear that    for a given intrinsic  burst duration larger $\Omega$  produce emission that is  seen for longer observer
time. This is due to the fact that   the line of sight samples larger angular range and 
correspondingly larger differences in the line 
 of sight advances of the emitting region.

\begin{figure}[h!]
  \centering
  \includegraphics[width=0.3\textwidth]{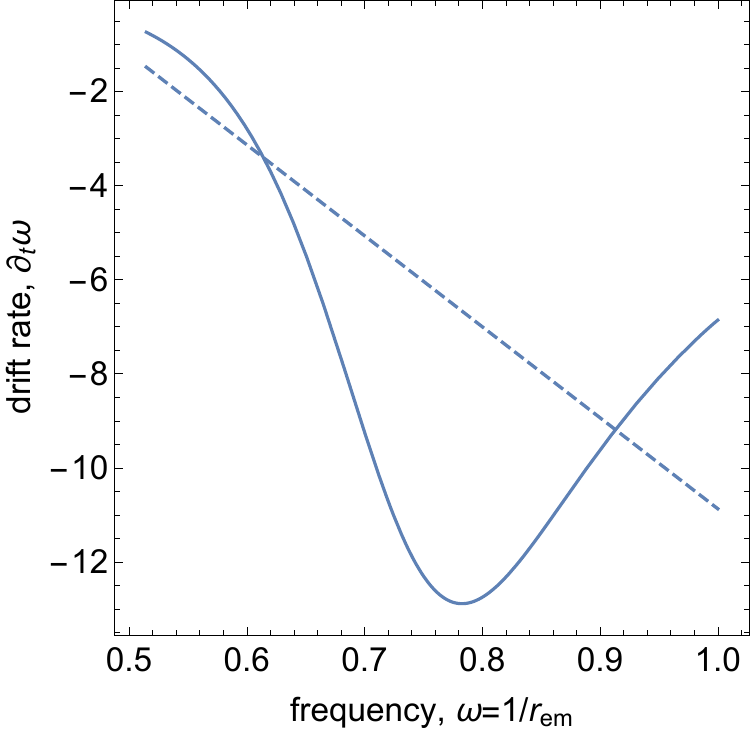}
  \includegraphics[width=0.3\textwidth]{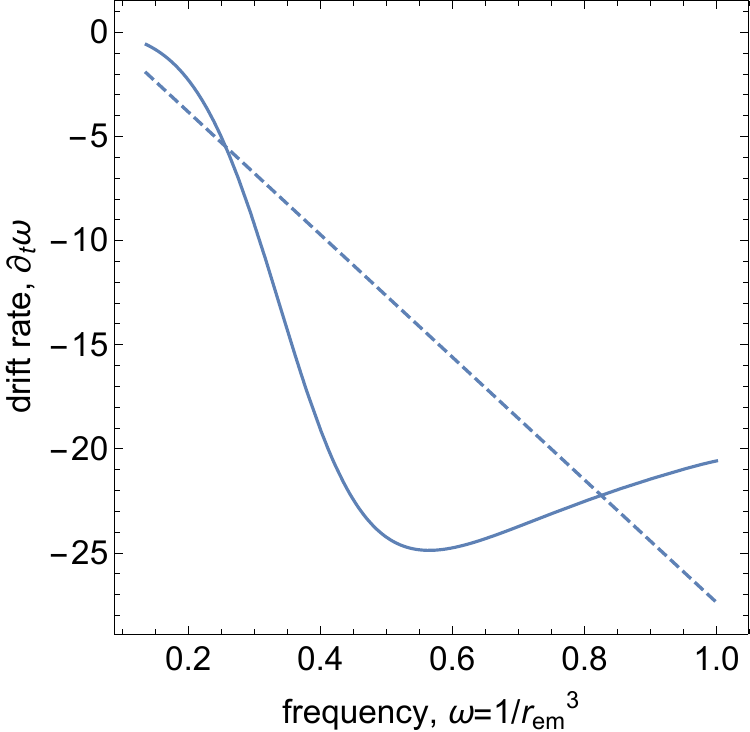}\\
  \includegraphics[width=0.3\textwidth]{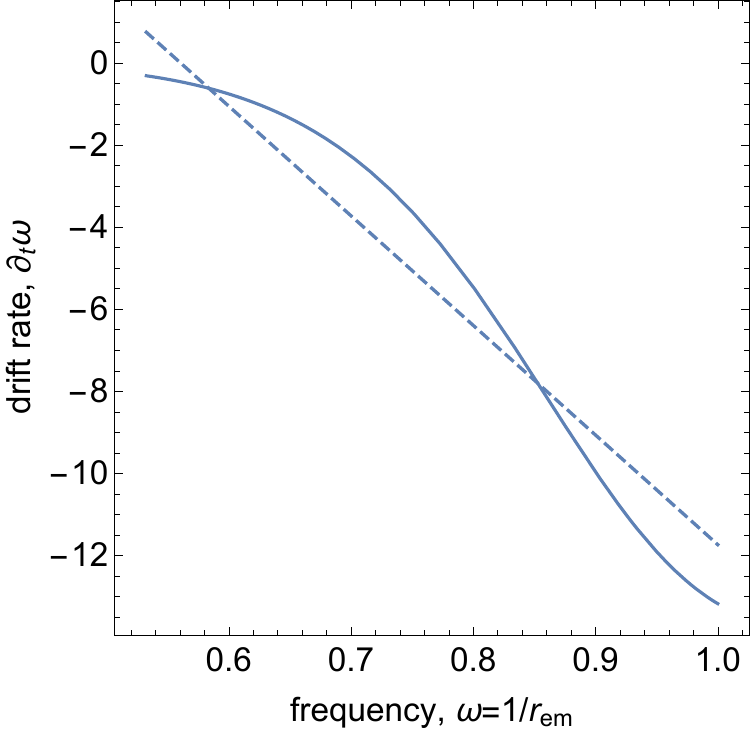}
  \includegraphics[width=0.3\textwidth]{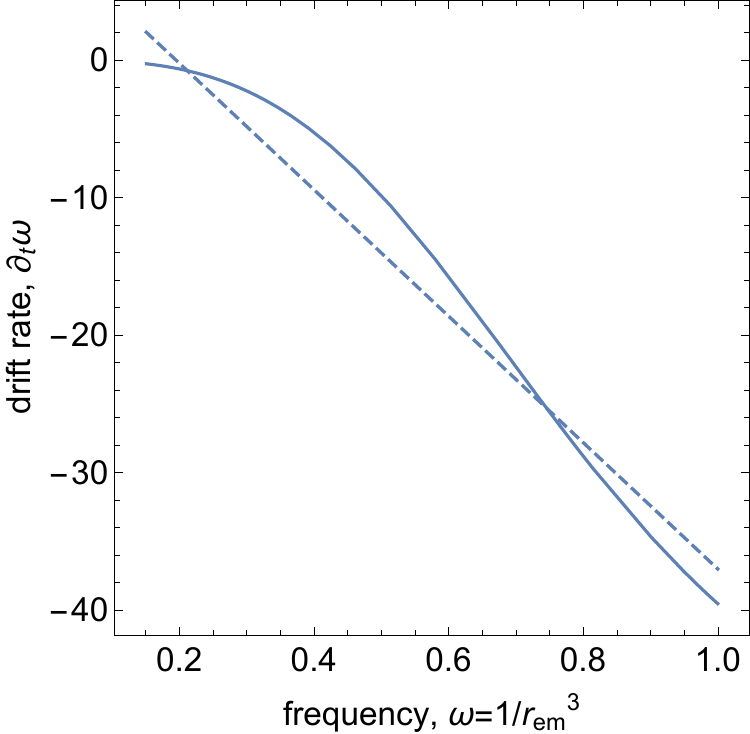}\\
   \includegraphics[width=0.3\textwidth]{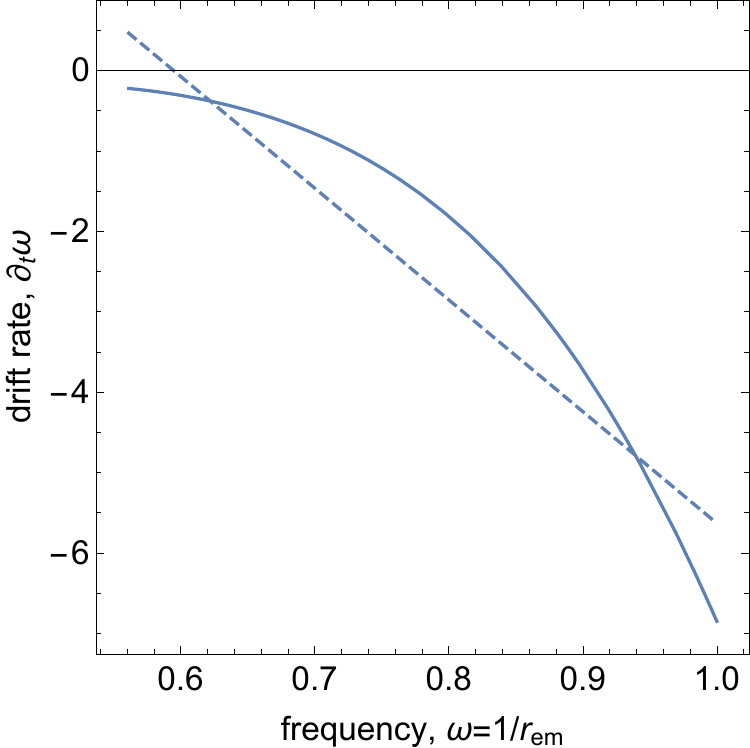}
  \includegraphics[width=0.3\textwidth]{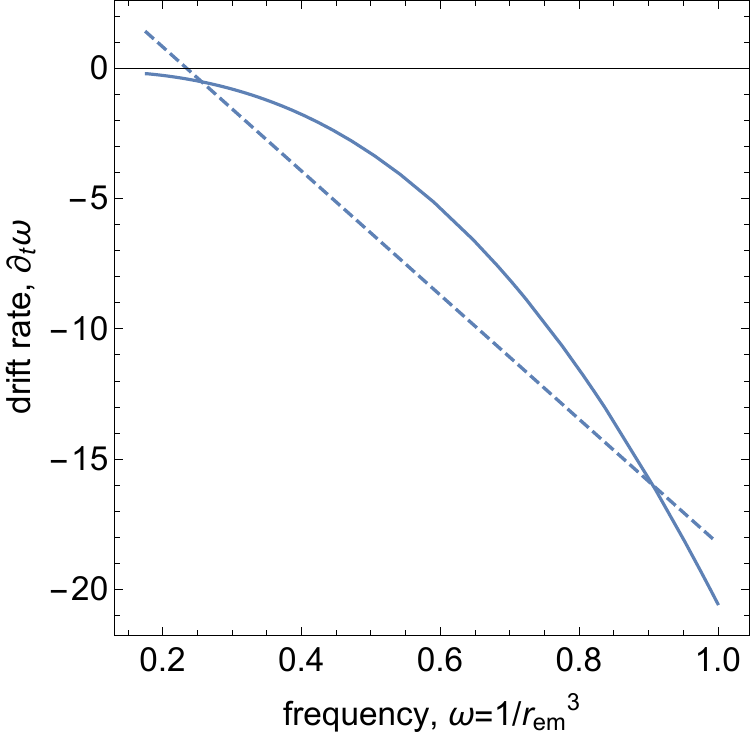}
  \caption{ Drift rates as function of frequency in rotating \mss; $\om \propto r_{em}^{-1}$ (left panels)  and  $\om \propto r_{em}^{-3}$ (right panels). Parameters are: 
   $\alpha=\pi/4$, $\theta _{ob}^{(0)} =\pi/2$, $\Omega = \pi2/$. Top row: $\Delta \phi = - \pi/4$, middle row:  $\Delta \phi =0$, bottom row:  $\Delta \phi = -\pi/4$.  Dashed lines are linear fits.  At intermediate frequencies the drifts rate are highly dependent on the spin frequencies, while at higher frequencies the  curves converge and hence less sensitivity to spin.  This example shows that the model can produce/predicts a variety of frequency drifts.}
     \label{drift2}
  %file FRB-drift
 \end{figure}

 In Fig. \ref{drift2} we show  the corresponding frequency drifts for selected parameters.  As is clear from the plots, the evolution of the peak frequency can be more complicated in the rotation \mss, as the line of sight samples large variations of plasma parameters. 
 (The dimensionless spin $\Omega = \pi/2$ in Fig. \ref{drift2}  is relatively high; smaller $\Omega$ produce more linear scalings)

   \begin{figure}[h!]
  \centering
  \includegraphics[width=0.3\textwidth]{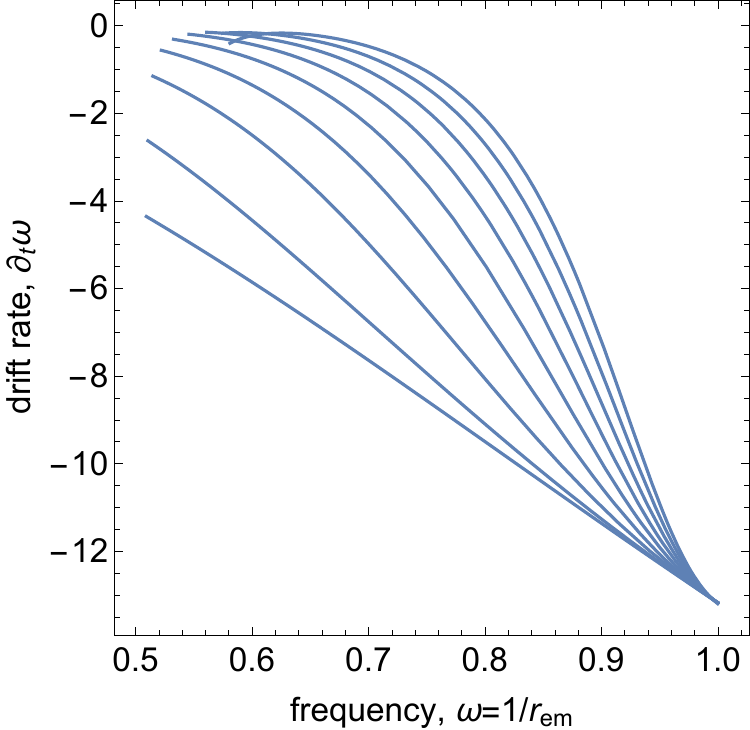}
  \includegraphics[width=0.3\textwidth]{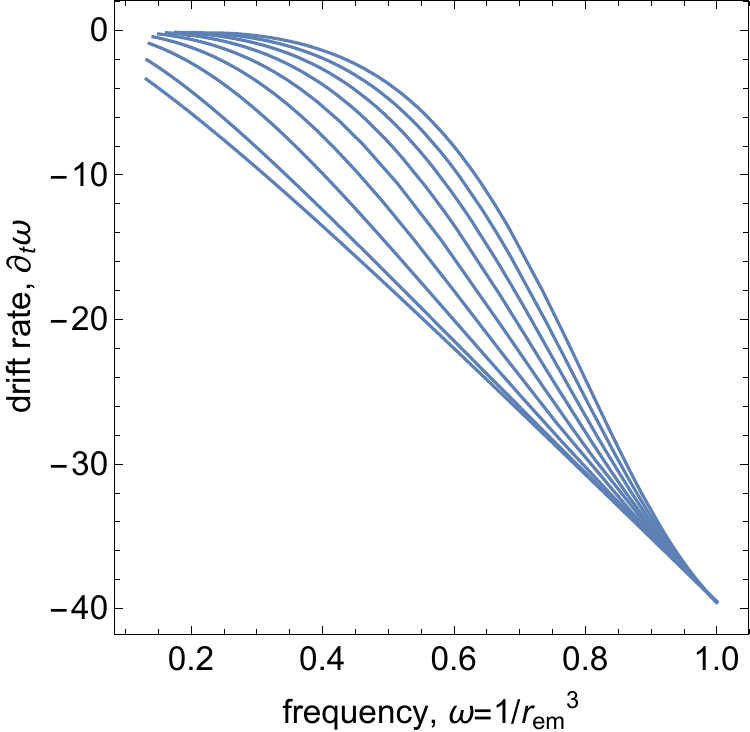}
  \caption{ Drift rates as function of frequency in rotating \mss; $\om \propto r_{em}^{-1}$ (left panel)  and  $\om \propto r_{em}^{-3}$ (right panel). Parameters are: 
   $\alpha=\pi/4$, $\theta _{ob}^{(0)} =\pi/2$,  $\Delta \phi=0$.  Different curves correspond to different spin frequencies  $\Omega =0,\, \pi/8... \pi$ (bottom to top).  Thus, the rotation of a   \NS\ does affect the frequency drifts.  Closer to $r_0$ (higher frequencies)  higher spins result in large frequency drifts. }
  \label{drift22}
  %file FRB-drift
 \end{figure}

\begin{figure}[h!]
  \centering
  \includegraphics[width=0.99\textwidth]{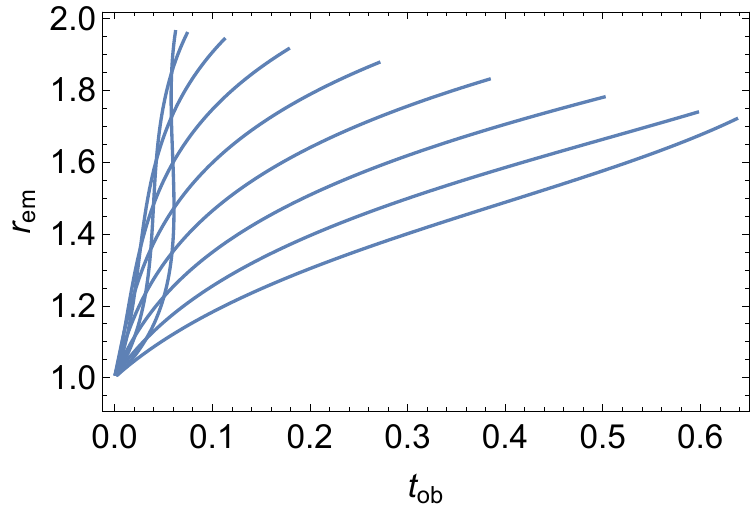}
  \caption{ Curves  $r_{em}(t_{ob}) $ for different launching  phases o $\Delta \phi = -\pi/2 -- \pi/2$ in steps of $\pi/8$ ( $\alpha=\pi/4$,   $\theta _{ob}^{(0)} =\pi/2$, $\Omega = \pi/2$. This demonstrates that different behavior can be seen from the same object depending on the initiation  moment of the emission front.}
  \label{drift14}
  %file FRB-drift
 \end{figure}

Importantly, 
depending on the trigger phase $\Delta \phi$ the same object will produce different $r_{em}(t_{ob}) $ curves, Fig. \ref{drift14}. This explains why in the Repeaters FRB 121102 different burst have different drifts \citep{2019ApJ...876L..23H}. The fact that different parts of the magnetar \ms\ can become active also explains the lack of periodicity in repeating FRBs.

\subsubsection{Prediction: polarization swings}

Polarization behavior of FBRs is, arguably, the most confusing overall \citep{2015Natur.528..523M,2015MNRAS.447..246P,2019MNRAS.487.1191C,2019A&ARv..27....4P}. We are not interested here in the propagation effects (\eg  sometimes huge and sometimes not RM measure). There is a clear, repeated detection of linear polarization. Importantly,  FRBs have thus
far not shown large polarization angle swings \citep{2019A&ARv..27....4P}.

The present model does not address the origin of polarization, as it would depend on the particular coherent emission mechanism. On basic grounds, polarization is likely to be determined by the local \Bf\ within the \ms. The model then
does predict polarization angle swings.
 In the rotating vector model \citep[RVM,][]{1969ApL.....3..225R} polarization swings reflect a local direction of the \Bf\ at the emission point. 
In our notations the position angle of polarization $\chi$  is given by 
 \be
 \tan \chi  =\frac{\sin\theta _{{ob}} ^{(0)} \sin ( \Omega t  + \Delta \phi) }{\cos \alpha 
   \sin\theta _{{ob}} ^{(0)}  \cos  ( \Omega t  + \Delta \phi) -\sin \alpha  \cos
  \theta _{{ob}}^{(0)}}
    \label{phi}
   \ee
     
\begin{figure}[h!]
  \centering
  \includegraphics[width=0.99\textwidth]{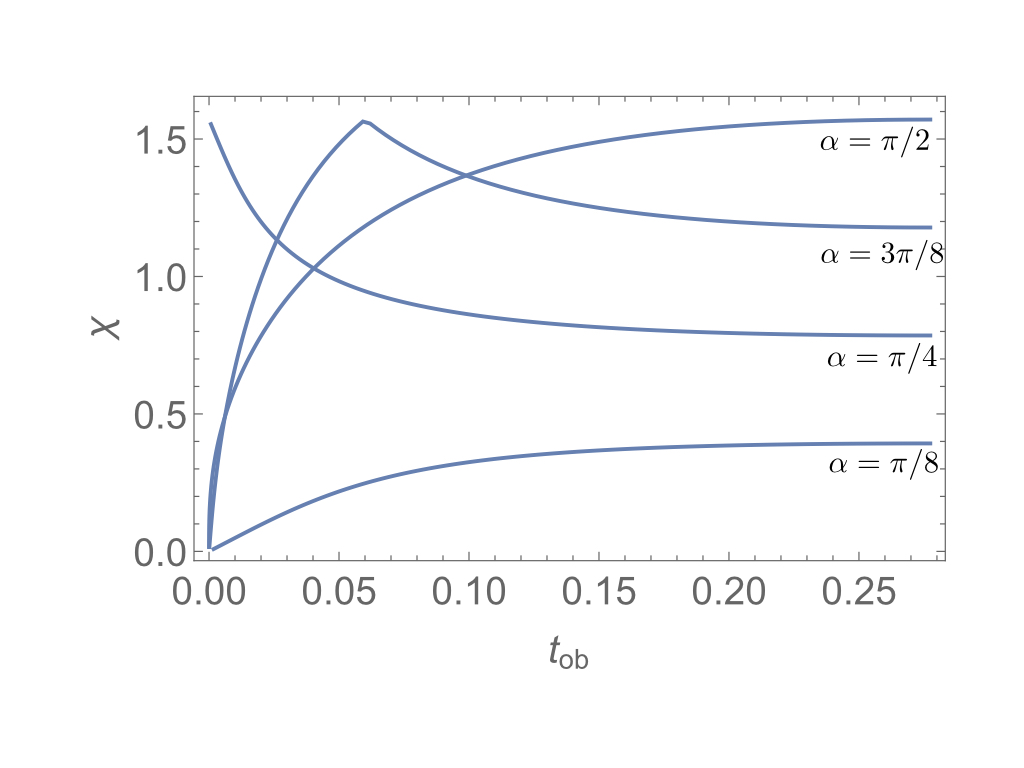}
  \caption{Position angle $\chi$ as function ob the observer time $t_{ob}$ for different $\alpha=0, ... \pi/2 $ in steps of $\pi/8$; $\Delta \phi=0$, $\Omega=\pi/2$, $\theta _{ob}^{(0)} =\pi/2$. }
  \label{chioftob}
  %file FRB-drift
 \end{figure}

Generally, we do expect polarization swings through the pulse,  Fig. \ref{chioftob}. Qualitatively, the fastest rate of change of the position angle occurs when the line of sight passes close to the magnetic axis; this requires $ \alpha  \approx \theta _{{ob}} ^{(0)}$ (so that the denominator comes close to zero). This is the case for rotationally powered pulsars. If emission is generated far from the magnetic axis, the expected PA swings are smaller. 
Thus, we do predict that PA swings will be observed within the pulses, but with values smaller than the ones seen in radio pulsars.

\section{Discussion}

In this paper we further argue that frequency drifts observed in FRBs point to the \mss\ of \NSs\ as the origin \citep[see also][]{2019arXiv190807313L}.  Our preferred model is a young  magnetar-type pulsar producing reconnection events during magnetic relaxation in the \mss\ \citep{2013arXiv1307.4924P}. 
 This should be a special type of magnetars, as there are observational constraints against radio bursts associated with the  known magnetars \citep[see, \eg\ discussion in ][]{2019arXiv190807313L}
 
 In astronomical setting, the repeater FRB 121102 is localized to an active star-forming galaxy, where ones does naturally  expects young \NSs\ \citep{2017ApJ...834L...7T}. 
 In contrast,  FRB 180924    is identified with galaxy   dominated by an old stellar
population with low star formation rate  \citep{2019Sci...365..565B}. One possibility is the formation of a \NS\ from an accretion induced collapse of a white dwarf   with a formation of a neutron star; this process  is probably responsible for formation of young pulsars in globular
clusters \citep{1996ApJ...460L..41L}.

The main points of the work are: 
\begin{itemize}
\item The observed  drift rate (\ref{drift00}) implies sizes of the order of \mss\ of \NSs. This is a somewhat independent constraint on the emission size, in addition to total duration of FRBs (which also is consistent with magnetospheric size). 
\item The observed  linear scaling of drift rate with frequency,   Eq.  (\ref{drift00}),   is a natural consequence of radius-to-frequency mapping in \mss\ of \NSs. It is valid, generally, for any power-law type  $\om(r_{em})$ dependence.   In fast rotating pulsars the drifts can have more complicated structure, \eg\ Fig. \ref{drift22}.
\item Non-observation of drifts in slowly rotating regular magnetars \citep[during  radio bursts][]{{2019ApJ...882L...9M}} is possibly due to the fact that higher spins may lead to higher drift rate, Fig. \ref{drift22}, higher amplitudes, Fig. \ref{FBR-drift}, and longer observer duration, Fig. \ref{remofTobOmega}.
\item In each given (repeating) FRB emission can originate at arbitrary rotational  phases, resulting in different drift profiles in different pulses from a given repeater, Fig. \ref{drift14}.
\end{itemize}

The model has a number of predictions.
\begin{itemize}
\item Polarization swings within the bursts,  reminiscent of  RVM for pulsars, can be observed. The amplitude of the swings in FRBs is expected to be smaller than in radio pulsar since the emission sights are not limited to the region near the magnetic axis, where PA  swings are the largest. 
\item For some parameters (line of sight, magnetic inclination and spin)  the frequency drifts are not linear  in frequency, \eg\ Fig. \ref{drift22}.   Given a limited signal to noise ratio of the typical data,   regular, continuous frequency drifts are easier identifiable; more complicated ones are more difficult to find during the  de-dispersion procedure. We encourage searchers for more complicated frequency drifts within FRBs. 

\end{itemize}

%Finally, let us comment on the possible observational breakthrough we foresee: (i)  simultaneous radio--X-ray observations of bursts from the local magnetars (but the corresponding FRBs-like events will be of much lower  radio luminosity); (ii)  ``shadowing'' simultaneous observations in radio and optical of the same field (and hopefully a simultaneous detection), as argued by \cite{2017ApJ...838L..13L}.

  %%%%%%%%%%%%%%%%%%%%%%%%%%%%%%%%%%%%%%%%%%%%%%%%%%%%%%%%%%%%%%%%%
\section*{Acknowledgments}
This work had been supported by DoE grant DE-SC0016369,
NASA grant 80NSSC17K0757 and  NSF grants  1903332 and 1908590. I would like to thank  Roger Blandford,  Jason Hessels, Victoria Kaspi and Amir Levinson  for discussions and comments on the manuscript.  

%%%%%%%%%%%%%%%%%%%%%%%%%%%%%%%%%%%%%%%%%%%%%%%%%%%%%%%%%%%%%%%%%

\bibliographystyle{apj}
  \bibliography{/Users/maxim/Home/Research/BibTex}

\end{document}